\documentclass[lettersize,journal]{IEEEtran}
\usepackage{amsmath,amsfonts}
\usepackage{algorithm,algpseudocode}
\usepackage{array}
\usepackage[caption=false,font=normalsize,labelfont=sf,textfont=sf]{subfig}
\usepackage{textcomp}
\usepackage{stfloats}
\usepackage{url}
\usepackage{verbatim}
\usepackage{graphicx}
\usepackage{cite}
\usepackage{cases}
\usepackage[dvipsnames]{xcolor} 
\usepackage[version=4]{mhchem} 
\usepackage{makecell} 

\begin{document}
\bstctlcite{IEEEexample:BSTcontrol} 

\title{A Stochastic Biofilm Disruption Model based on Quorum Sensing Mimickers}

\author{Fatih Gulec,~\IEEEmembership{Member,~IEEE,} Andrew W. Eckford,~\IEEEmembership{Senior Member,~IEEE}  
\thanks{The authors are with the Department of EECS, York University, Toronto M3J 1P3, Canada (Email: \{fgulec, aeckford\} @yorku.ca)}       
\thanks{The works of FG and AWE were funded by a Discovery grant from the Natural Sciences and Engineering Research Council of Canada.}
\thanks{This work was presented at the $7^{th}$ Workshop on Molecular Communications which was held in Erlangen, Germany and accepted for publication in IEEE Transactions on Molecular, Biological, and Multi-Scale Communications.}%
}



\maketitle

\begin{abstract}
Quorum sensing (QS) mimickers can be used as an effective tool to disrupt biofilms which consist of communicating bacteria and extracellular polymeric substances (EPS). In this paper, a stochastic biofilm disruption model based on the usage of QS mimickers is proposed. A chemical reaction network (CRN) involving four different states is employed to model the biological processes during the biofilm formation and its disruption via QS mimickers. In addition, a state-based stochastic simulation algorithm is proposed to simulate this CRN. The proposed model is validated by the \textit{in vitro} experimental results of \textit{Pseudomonas aeruginosa} biofilm and its disruption by rosmarinic acid as the QS mimicker. Our results show that there is an uncertainty in state transitions due to the effect of the randomness in the CRN. In addition to the QS activation threshold, the presented work demonstrates that there are underlying two more  thresholds for the disruption of EPS and bacteria, which provides a realistic modeling for biofilm disruption with QS mimickers.
\end{abstract}


\section{Introduction}
\IEEEPARstart{B}{iofilms} can be defined as bacterial cities where communicating bacteria live together. A biofilm mainly consists of a bacterial population and extracellular polymeric substances (EPS). They are related to negative effects on human health, since they can cause infection or antibiotic resistance. Therefore, modeling the disruption of the biofilm is essential.

During the growth phase of the biofilm, bacteria use a cell-to-cell communication mechanism called quorum sensing (QS). In QS, they send to each other autoinducer molecules to control if their population is sufficient in the medium. Once they sense that they reach a sufficient population, they trigger intracellular mechanisms such as aggregation, biofilm formation and production of virulence factors \cite{perez2016mathematical}. When the biofilm is formed on a surface, it is more difficult to eradicate it with chemical and mechanical disruption methods \cite{paluch2020prevention}. Hence, methods to exploit the QS mechanism are investigated to prevent or disrupt biofilm formation in the literature.

To this end, quorum quenching (QQ) strategy which includes methods to inhibit the communication among bacteria is proposed \cite{paluch2020prevention}. These QQ methods include the degradation and inhibition of autoinducer molecules and blocking the autoinducer reception via blocking the intracellular signal transduction pathways. Another strategy based on exploiting the QS mechanism is the usage of QS mimickers, which can bind to autoinducer receptors of bacteria, employed in inter-kingdom signaling \cite{papenfort2016quorum}. For example, these QS mimickers are employed as a defense mechanism in plants to disrupt the biofilm with an early induction of QS. Rosmarinic acid, which is a QS mimicker secreted as a plant defense compound, triggers an early QS activation and then can kill the bacteria and eradicate the EPS in the biofilm \cite{corral2016rosmarinic}. In this study, it is shown that rosmarinic acid can bind to the autoinducer receptors of bacteria and can compete with homoserine lactone molecules which are produced and processed by \textit{Pseudomonas aeruginosa} type bacteria for QS.

As for the modeling of biofilm disruption via exploiting the QS, several methods in systems biology and molecular/biological communication literature are proposed. In \cite{fozard2012inhibition}, an individual based model is proposed to observe the effect of QS inhibition in biofilm formation. In \cite{martins2016using}, a deterministic bacterial wall model consisting of communicating bacteria is proposed to disrupt the bacteria via starvation. In \cite{martins2018molecular}, a deterministic biofilm suppression model in which QS signals are jammed is proposed. In addition, QS in a bacterial community is modeled by using a queuing model in \cite{michelusi2016queuing}. However, none of these models focuses on the disruption of the biofilm, i.e., killing bacteria and eradicating EPS, by exploiting the QS mimicking mechanism.

In this paper, a stochastic biofilm disruption model by using QS mimickers is proposed. The biological phenomena for the formation and disruption of the biofilm are modeled via coupled chemical reactions, i.e., a chemical reaction network (CRN). This model is based on four biological states. The first two states represent the formation of the biofilm before (downregulation) and after (upregulation) the QS activation according to the autoinducer and QS mimicker concentrations. In these states, QS mimickers help for an earlier QS response. In the last two states, disruption is modeled via two different thresholds. Depending on the QS mimicker concentration, firstly, EPS is eradicated and then bacteria are killed in the last state. Furthermore, a state-based stochastic simulation algorithm is proposed to simulate the CRN using these biological states. Our results are validated with the  experimental results of \textit{Pseudomonas aeruginosa} type as bacteria and rosmarinic acid as QS mimicker. Our model is able to show the stochasticity in the transition of the biological states and stochastic changes in the bacterial and EPS concentrations within the biofilm due to the randomness of the chemical reactions in the medium. The main contributions of this paper are to provide a state-based CRN model for the disruption of the biofilm via QS mimickers and a state-based stochastic simulation algorithm.

\section{Model}
\begin{figure}[h]
	\centering
	\includegraphics[width=1.00\columnwidth]{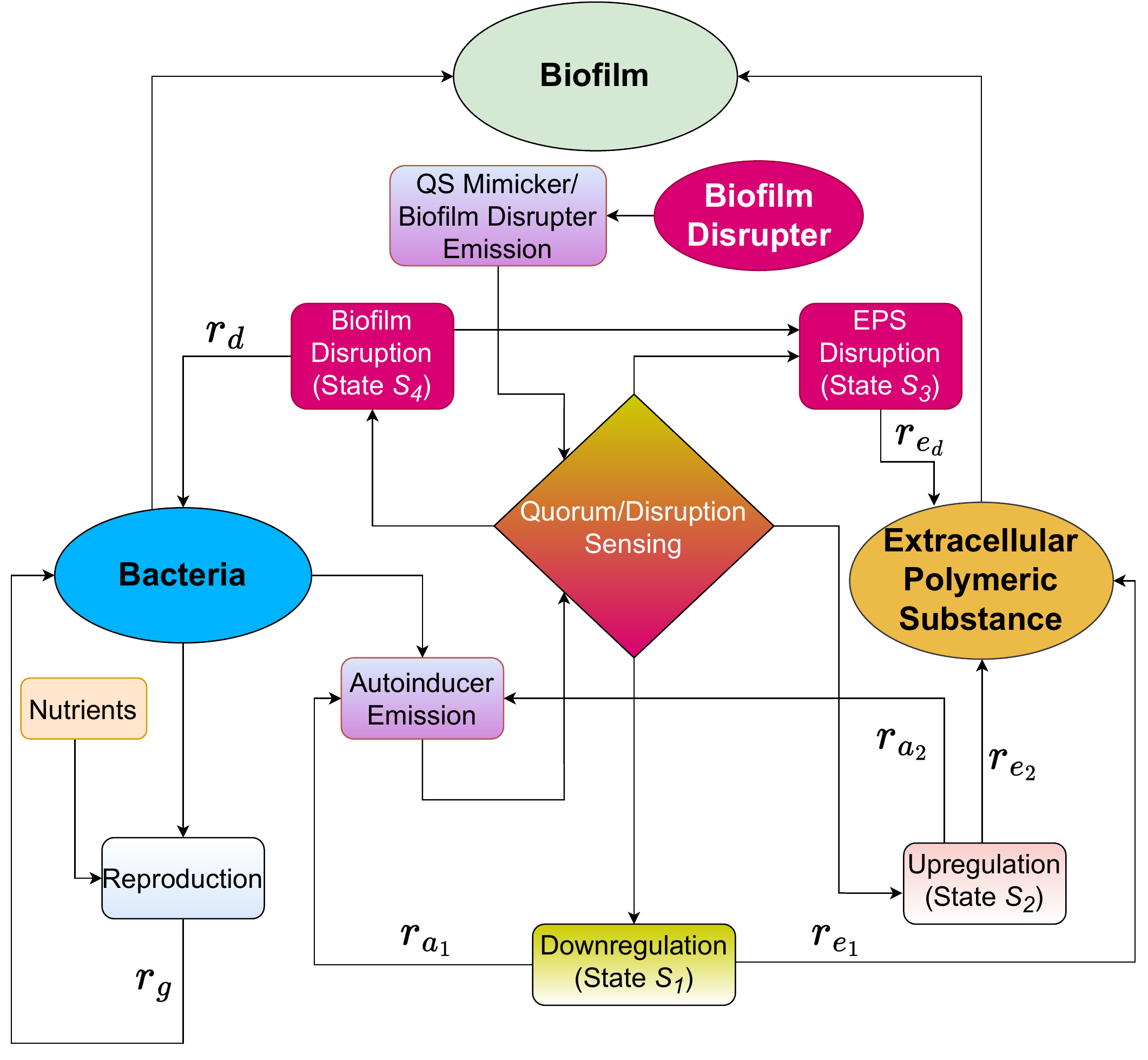} 
	\vspace{-0.6cm}
	\caption{
		Biological processes/states for biofilm formation and its disruption. The symbols over the arrows represent the stochastic reaction constants which are explained in Section \ref{CRN}.
	}
	\label{Biofilm_block}
		\vspace{-0.4cm}
\end{figure}

In this section, the biological processes related to biofilm formation and its disruption by the QS mimickers are detailed. Then, our proposed model is given to explain these biological processes based on Fig. \ref{Biofilm_block}.

\subsection{Biological Background of Biofilm Formation/Disruption}
The first stage in biofilm formation is the attachment of bacteria to a surface. After this attachment, they start to form a biofilm via the bacterial reproduction and EPS production. During this growth stage, bacteria use QS for communication among them by emitting autoinducer ($A$) molecules and thus induce intracellular mechanisms \cite{klapper2010mathematical}. Subsequent to the growth and maturation stage, biofilm disperses some of its bacteria to form new biofilms in the vicinity. In this paper, the biofilm formation during the growth stage is of interest for modeling.

In biofilm formation, QS mechanism can be employed to increase the EPS production rate. When the autoinducer concentration exceeds a threshold, bacteria pass from downregulation to upregulation state for a higher rate EPS production \cite{frederick2011mathematical}. Furthermore, biofilms can be disrupted by emitting biofilm disrupter molecules such as rosmarinic acid which is secreted by plants and acts also as a QS mimicker ($M$) \cite{corral2016rosmarinic}. These $M$ molecules trigger an earlier QS upregulation state. However, they also disrupt the biofilm in two different stages as shown by the \textit{in vitro} results in \cite{corral2016rosmarinic}. First, EPS is removed, when the concentration of $M$ is above a threshold. Second, bacteria begin to be killed in addition to the EPS disruption after the mimicker concentration exceeds a second disruption threshold. Hence, there are four states where two states are related to QS as downregulation and upregulation, and other two states define the EPS and biofilm (bacteria and EPS) disruption. Next, the biological processes given in this section are modeled based on these four states.



\vspace{-0.3cm}
\subsection{Chemical Reaction Network} \label{CRN}
In this paper, our first aim is to model the phenomena about the effect of QS mimickers on QS and biofilm disruption by using the \textit{in vitro} experimental results in \cite{corral2016rosmarinic}. Therefore, the QS-based biofilm formation and QS mimicker-based biofilm disruption are assumed to occur in the same homogeneous volume ($V$) as in \cite{corral2016rosmarinic}. In addition, rosmarinic acid which is a QS mimicker molecule is treated by bacteria as autoinducer molecules as shown with experimental results in \cite{corral2016rosmarinic}. Firstly, we assume that bacteria reproduce according to the availability of nutrients in the medium. All the aforementioned processes as summarized in Fig. \ref{Biofilm_block} are modeled as coupled chemical reactions, i.e., a CRN, based on four states ($S_1$-$S_4$). In this CRN given in (\ref{R1a})-(\ref{R9}) according to their states, the chemical species $ A $, $ B $, $ E $, $ S_\mathrm{N} $, $ M $ and $ C $ represent autoinducer molecules, bacteria, EPS, nutrient substrates, QS mimickers and nutrient-bacterium complex, respectively. In addition, $\emptyset$ shows the species that are of no interest and $Y_{b/s}$ is the yield coefficient for nutrient consumption. Moreover, it should be noted that the variables on the arrows in (\ref{R1a})-(\ref{R9}) represent the stochastic reaction constants ($r_{a_1}$, $ r_{e_1} $, etc.) which are not always the same with deterministic reaction rate constants \cite{gillespie1976general}.

\paragraph*{\underline{State $S_1$}} 
\begin{center}	
	\vspace{-0.5cm}
\begin{tabular} {p{0.4\columnwidth}p{0.45\columnwidth}}
	\begin{equation} \label{R1a}
		\ce{\emptyset{} + B ->[r_{a_1}] A + B}
	\end{equation} &	
	\begin{equation} \label{R2a}
		\ce{\emptyset{} + B ->[r_{e_1}] B + E} 		
	\end{equation}
\end{tabular}
\end{center}
\vspace{-0.5cm}

\paragraph*{\underline{State $S_2$}} ~\\
\begin{center}
	\vspace{-0.8cm}	
\begin{tabular} {p{0.4\columnwidth}p{0.45\columnwidth}}
	\begin{equation}\label{R1b}
		\ce{\emptyset{} + B ->[r_{a_2}] A + B}
	\end{equation} &
	
	\begin{equation} \label{R2b}
		\ce{\emptyset{} + B ->[r_{e_2}] B + E} 		
	\end{equation}
\end{tabular}
\end{center}

\vspace{-0.5cm}
\paragraph*{\underline{State $S_3$}} 
\begin{equation} \label{R3}
	\ce{E ->[r_{e_d}] \emptyset{} }
\end{equation}

\vspace{-0.2cm}
\paragraph*{\underline{State $S_4$}} 
\begin{equation} \label{R4}
	\ce{B ->[r_{d}] \emptyset{}}
\end{equation}

\vspace{-0.2cm}
\paragraph*{\underline{States $S_1 - S_4$}} ~\\
\begin{center}
	\vspace{-0.4cm}	
	\begin{tabular} {p{0.4\columnwidth}p{0.45\columnwidth}}
		\begin{equation} \label{R5}
			\ce{\emptyset{} ->[r_m] M} 		
		\end{equation}&		
		\begin{equation} \label{R6}
			\ce{A ->[r_\sigma] \emptyset{}} 		
		\end{equation}
\end{tabular}
\end{center}

\begin{center}
	\vspace{-0.8cm}	
	\begin{tabular} {p{0.4\columnwidth}p{0.45\columnwidth}}
		\begin{equation} \label{R7}
			\ce{B + S_N ->[r_c] C} 		
		\end{equation} &
		
		\begin{equation} \label{R8}
			\ce{C ->[r_g] (1 + Y_{b/s})B} 		
		\end{equation}
\end{tabular}
\end{center}

\vspace{-0.5cm}
\begin{equation} \label{R9}
	\ce{M ->[r_{dm}] \emptyset{}} 		
\end{equation}

In this CRN, reactions (\ref{R5})-(\ref{R9}) occur in all states. Reactions (\ref{R5}) and (\ref{R9}) represent the production and degradation of $M$, respectively. Autoinducer molecules degrade with the rate $r_\sigma$ as given in (\ref{R6}). Reactions (\ref{R7}) and (\ref{R8}) show the bacterial growth based on Monod kinetics in a chemical reaction form \cite{alvarez2019theoretical}. In (\ref{R7}), bacteria consume nutrient substrates with a rate $r_c$ to produce $C$ which actually represents the bacteria consuming the nutrients. Then, these $C$ complexes produce new bacteria with the rate $r_g$ as given in (\ref{R8}). $r_g$ can be calculated as \cite{alvarez2019theoretical}
\begin{equation} \label{r_g}
	r_g = \frac{\mu_\mathrm{max}(1+Y_{b/s})}{Y_{b/s}},
\end{equation}
where $\mu_\mathrm{max}$ is the maximum specific growth rate (h$^{-1}$). Besides, the stochastic production constant for $C$ can be given as $r_c = k_c C_g$ where $C_g$ is the nutrient concentration (g/l) and $k_c$ is the deterministic reaction rate constant as given by \cite{alvarez2019theoretical}
\begin{equation}
	k_c = \frac{r_g}{(1+Y_{b/s}) K_M},
\end{equation}
where $K_M$ is the Monod constant (g/l). 

Furthermore, reactions (\ref{R1a}) and (\ref{R2a}) show the productions of $A$ and $E$ with low rates ($r_{a_1}$ and $r_{e_1}$) at state $S_1$, respectively. In state $S_2$, (\ref{R1b}) and (\ref{R2b}) represent the production of $A$ and $E$ in higher rates ($r_{a_2}$ and $r_{e_2}$), respectively. In state $S_3$, $E$ is disrupted with the rate $r_{e_d}$ as shown in (\ref{R3}). In state $S_4$, bacteria are disrupted with the rate $r_d$ according to (\ref{R4}), while the EPS disruption in this state continues according to (\ref{R3}). The state decisions are made according to the detection rules given in (\ref{S1})-(\ref{S4}) by using the concentrations of $A$ ($C_A(t)$) and $M$ ($C_M(t)$), QS detection threshold ($\Gamma_\mathrm{QS}$), EPS disruption threshold ($\Gamma_\mathrm{DE}$) and biofilm (bacteria and EPS) disruption threshold ($\Gamma_\mathrm{DB}$) with the condition $\Gamma_\mathrm{QS}<\Gamma_\mathrm{DE}<\Gamma_\mathrm{DB}$. 

\vspace{0.1cm}

\paragraph*{\underline{State $S_1$ - QS Downregulation}}
\begin{equation} \label{S1}
	C_M(t) < \Gamma_\mathrm{DE} \quad \text{and} \quad C_A(t) + C_M(t) < \Gamma_\mathrm{QS}
\end{equation}

\paragraph*{\underline{State $S_2$ - QS Upregulation}}
\begin{equation} \label{S2}
	C_M(t) < \Gamma_\mathrm{DE} \quad \text{and} \quad C_A(t) + C_M(t) \geq \Gamma_\mathrm{QS}
\end{equation}

\paragraph*{\underline{State $S_3$ - EPS Disruption}}
\begin{equation} \label{S3}
	\Gamma_\mathrm{DE} \leq C_M(t) < \Gamma_\mathrm{DB}
\end{equation}

\paragraph*{\underline{State $S_4$ - Biofilm Disruption}}
\begin{equation} \label{S4}
	C_M(t) \geq \Gamma_\mathrm{DB}.
\end{equation}

When the sum of autoinducer and QS mimicker concentration exceeds the threshold $\Gamma_\mathrm{QS}$, then they pass to a upregulation state (state $S_2$). As shown in (\ref{S1}) and (\ref{S2}), states $S_1$ and $S_2$ are active when $C_M(t)$ is below the EPS disruption threshold $\Gamma_\mathrm{DE}$. Once the concentration of $M$ reaches the EPS disruption threshold $\Gamma_\mathrm{DE}$, the EPS disruption state (state $S_3$) is activated. In this state, only EPS is disrupted. In the last state (state $S_4$), bacteria start to be disrupted in addition to EPS after the biofilm disruption threshold $\Gamma_\mathrm{DB}$ is reached. 

Next, these state-based decisions are employed for the stochastic simulation of the CRN. 

\section{State-based Stochastic Simulation Algorithm} \vspace{-0.0cm}
In this section, the stochastic simulation method based on the biological states given in the previous section is elaborated. As for the analytical stochastic method, chemical reactions can be characterized via chemical master equations (CMEs). These CMEs are employed to determine the probability that each species in a CRN have a certain number of particles. However, they are not feasible to use for realistic CRNs with a high number of reactions as our case in this paper. Therefore, stochastic simulation algorithms are used to simulate the time course of the number of particles for each species \cite{gulec2022stochastic}. In this work, the direct Gillespie algorithm which is an exact stochastic simulation method is employed \cite{gillespie1977exact}. However, since the biological states change the reaction probabilities in this method, we propose a state-based stochastic simulation algorithm (SbSSA) based on the direct Gillespie algorithm detailed as follows.

Let $\mathbf{X}(t) = \left( A(t), B(t), E(t), M(t), S(t), C(t) \right)$ be the vector which holds the number of particles for all species in the CRN. The change vector which shows the change in the number of species with respect to the stoichiometric coefficients in (\ref{R1a})-(\ref{R9}) is also defined as $\pmb{\nu}_j = (\nu_{j,1},...,\nu_{j,6})$ for the $j^{th}$ reaction. Furthermore, the probability for the $ j^{th} $ reaction to occur in the infinitesimal period $[t, t+dt)$ is given by $a_j(\mathbf{x}) dt$ where $a_j(\mathbf{x})$ is the propensity function showing the probability per unit time of the given reaction when $\mathbf{X}(t) = \mathbf{x}$. Lastly, we define the propensity vector for the CRN as $\mathbf{a} = (a_1, a_2,...,a_9)$. Here, this vector is defined for nine reactions, since reactions (\ref{R1a})-(\ref{R2a}) represented by $a_1$ and reactions (\ref{R1b})-(\ref{R2b}) represented by $a_2$ cannot occur simultaneously. Thus, the first four elements of the propensity vector are based on the biological states ($S_1$-$S_4$) and they are updated according to Algorithm \ref{Alg1}. Here, $X(1)$, $X(2)$, $X(3)$, and $X(4)$ correspond to the numbers of autoinducer, bacteria, EPS and QS mimicker molecules.

\setlength{\textfloatsep}{1pt}

\begin{algorithm}[tb]
	\caption{State-based Stochastic Simulation Algorithm}
	\label{Alg1} 
	\begin{algorithmic}[1]
		\While{$ t \leq t_s $}
			\If{$C_M(t) < \Gamma_\mathrm{DE}$ \textbf{and} $C_A(t) + C_M(t) < \Gamma_\mathrm{QS}$}
			\State $a(1:4) \gets  [r_{a_1} X(2), r_{e_1} X(2), 0, 0 ] $ \Comment{$S_1$}
			\ElsIf{$ C_M(t) \hspace{-0.1cm} < \hspace{-0.1cm} \Gamma_\mathrm{DE} $ \textbf{and} $ C_A(t) + C_M(t) \hspace{-0.1cm} \geq \hspace{-0.1cm} \Gamma_\mathrm{QS} $}	
			\State $a(1:4) \gets  [r_{a_2} X(2), r_{e_2} X(2), 0, 0 ] $ \Comment{$S_2$}
			\ElsIf{$ \Gamma_\mathrm{DE} \leq C_M(t) < \Gamma_\mathrm{DB} $}
			\State $a(1:4) \gets  [0, 0, r_{e_d} X(3), 0] $ \Comment{$S_3$}
			\Else	
			\State $a(1:4) \gets  [0, 0, r_{e_d} X(3), r_d X(2)] $ \Comment{$S_4$}		
			\EndIf
			\State Update state-independent elements of $\mathbf{a}$ \Comment{All states}		
			\State Determine $j$ and $\tau$ via Gillespie algorithm \cite{gillespie1977exact}
			\State $\mathbf{X} \gets \mathbf{X} + \pmb{\nu}_j$
			\State $t \gets t + \tau$
			\State $C_A(t) \gets X(1)/V$; $C_M(t) \gets X(4)/V$
		\EndWhile	
	\end{algorithmic} 
\end{algorithm}

\setlength{\textfloatsep}{20.0pt plus 2.0pt minus 4.0pt} 

\begin{table}[b]
	\centering
	\caption{Simulation parameters}
	\centering\setcellgapes{2pt}\makegapedcells \renewcommand\theadfont{\normalsize\bfseries} 
	\scalebox{1}{
		\begin{tabular}{p{32pt}|p{65pt}|p{32pt}|p{65pt}}
			\hline
			\textbf{Parameter}	& \textbf{Value} & \textbf{Parameter}	& \textbf{Value}\\
			\hline \hline
			$r_{a_1}$ & $7.6$ nmol h$^{-1}$ \cite{henkel2013kinetic} & $r_{a_2}$ & $21.8$ nmol h$^{-1}$ \cite{henkel2013kinetic} \\
			$r_{e_1}$ & $0.035$ h$^{-1}$ \cite{frederick2011mathematical} & $r_{e_2}$ & $0.35$ h$^{-1}$ \cite{frederick2011mathematical} \\
			$r_{e_d}$ & $0.35$ h$^{-1}$ & 	$r_{d}$ & $7.5178$ h$^{-1}$ \\
			$r_{m}$ & $872$ nmol h$^{-1}$ & $r_{\sigma}$ & $3.1$ nmol h$^{-1}$ \cite{henkel2013kinetic} \\
			$\mu_\mathrm{max}$ & $0.29$ h$^{-1}$ \cite{beyenal2003double} & $C_g$ & $0.005$ g l$^{-1}$ \cite{beyenal2003double} \\
			$r_{c}$ & $0.0858$ h$^{-1}$ & $r_{g}$ & $0.7518$ h$^{-1}$ \\
			$r_{dm}$ & $0.0031$ nmol h$^{-1}$ & $Y_{b/s}$ & $0.628$ \cite{beyenal2003double} \\
			$V$	& $0.02$ l \cite{corral2016rosmarinic} & $K_M$ & $0.0269$ g l$^{-1}$ \cite{beyenal2003double} \\
			$\Gamma_\mathrm{QS}$ & $50$ $\mu$mol l$^{-1}$ & $\Gamma_\mathrm{DE}$ & $2$ mmol l$^{-1}$ \cite{corral2016rosmarinic} \\
			$\Gamma_\mathrm{DB}$ & $7.8$ mmol l$^{-1}$ \cite{corral2016rosmarinic} & 	&	 \\
			\hline \hline           
		\end{tabular}
	}
	\label{Sim_parameters}
\end{table}

For a given simulation time ($t_s$) and thresholds, the state-dependent elements of $\mathbf{a}$ are determined according to the biological states as defined in (\ref{S1})-(\ref{S4}). Then, the elements of $\mathbf{a}$ which are not dependent on the states but only to the number of changing particles are updated according to the order of the corresponding reaction as defined in \cite{gillespie1976general}. Next, one reaction ($j$) to occur is chosen for each step and the corresponding time step ($\tau$) is determined by the Gillespie algorithm \cite{gillespie1977exact}. In Gillespie algorithm, the time step is calculated as $\tau = (1/a_0) \ln(1/r_1)$ where $a_0$ is the sum of all elements in $\mathbf{a}$ and $r_1$ is a random variable drawn from a uniform distribution between $0$ and $1$, i.e., $U(0,1)$. In addition, $j$ is chosen so that $\sum_{k=1}^{j-1} a_k < r_2 a_0 \leq \sum_{k=1}^{j} a_k$ where $r_2 \sim U(0,1)$. The number of molecules are updated via the addition of the $j^{th}$ change vector with $\mathbf{X}$ and time ($t$) is updated by increasing it with $\tau$. 

In Algorithm \ref{Alg1}, the well-known Gillespie algorithm is implemented. However, it can be replaced by other modified versions such as explicit or implicit tau-leap methods \cite{gillespie2007stochastic}. The main idea of Algorithm \ref{Alg1} is to change the propensities according to the biological states of the CRN whose time course is shown with the numerical results in the next section.

\begin{figure}[tb] 
	\centering
	\includegraphics[width=\columnwidth]{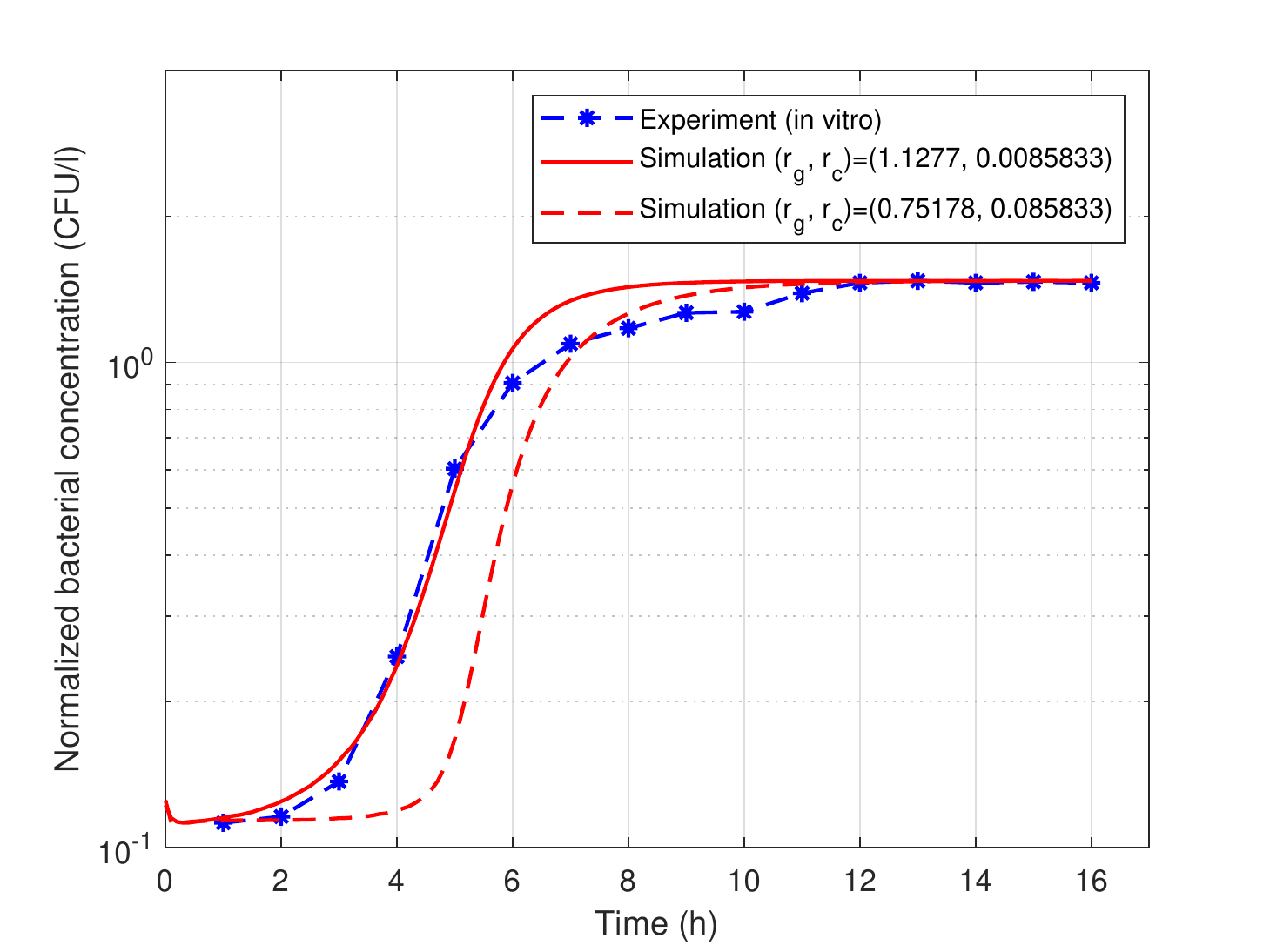}
			\vspace{-0.3cm}  
	\caption{Mean bacterial growth (in colony forming units (CFU) per liter).}
	\label{Plot_val_B}
\end{figure}

\begin{figure}[tb] 
	\centering
	\includegraphics[width=\columnwidth]{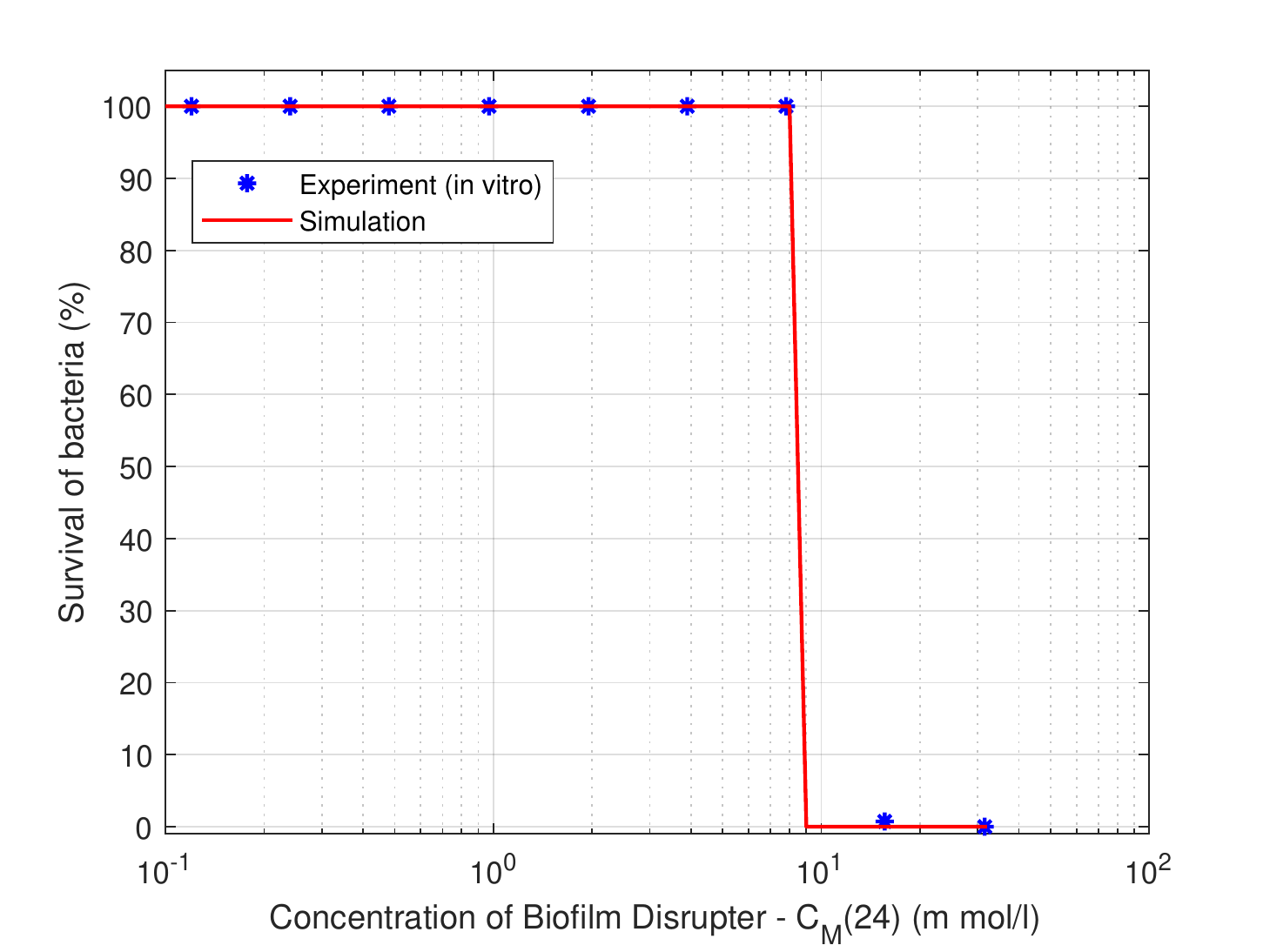}
		\vspace{-0.3cm} 
	\caption{Survival percentage of bacteria at the end of 24 h according to the QS mimicker concentration.}
	\label{Plot_val_dis}
\end{figure}

\begin{figure}[bt] 
	\centering
	\includegraphics[width=\columnwidth]{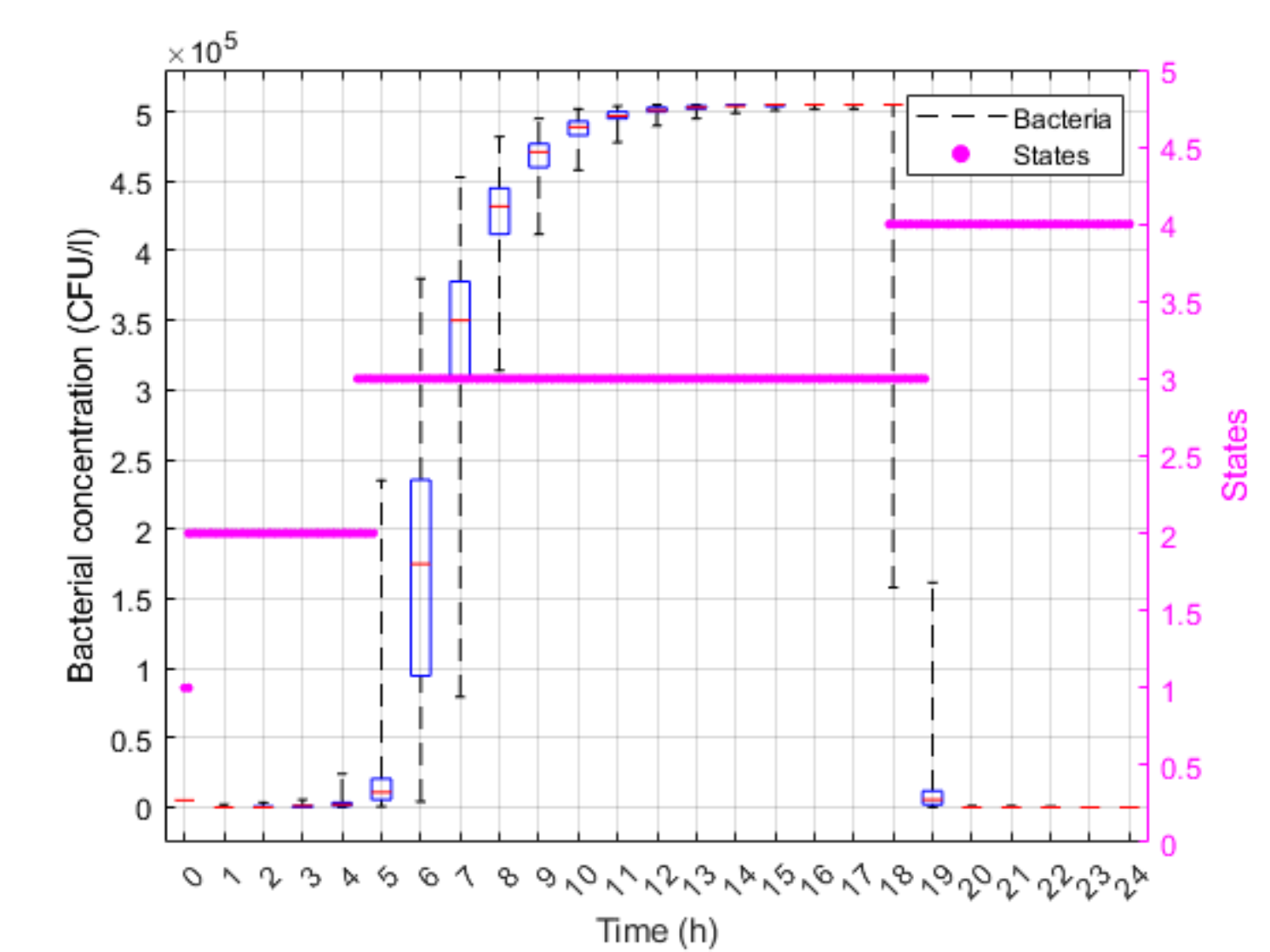}
	\caption{Distribution of states and box plot of bacterial concentration.}
	\label{Plot_biofilm_B}
\end{figure}

\section{Numerical Results}
In this section, numerical results which include the simulation and the validation with the \textit{in vitro} experimental results are given. As shown in Table \ref{Sim_parameters}, simulation parameters are mostly obtained from experimental works in \cite{henkel2013kinetic}, \cite{frederick2011mathematical}, \cite{beyenal2003double}, and \cite{corral2016rosmarinic} for \textit{Pseudomonas aeruginosa} type bacteria and rosmarinic acid as the QS mimicker. $r_c$ and $r_g$ are calculated by using the Monod kinetics by using $\mu_{max}$, $Y_{b/s}$, $K_M$ and $C_g$ values as  explained in Section \ref{CRN}. Since stochastic reaction constants of $A$ and $M$ are much higher than the other species in the CRN, it results in unfeasible simulation times. Therefore, we assume these species in units consisting of $1$ n mol particles and use these units in simulations to scale the related stochastic reaction constants, i.e., $r_{a_1}$, $r_{a_2}$, $r_{m}$, $r_{\sigma}$ and the corresponding thresholds, i.e., $\Gamma_\mathrm{QS}$, $\Gamma_\mathrm{DE}$ and $\Gamma_\mathrm{DB}$, which are only employed to determine the states.

Our proposed model is validated by the \textit{in vitro} experimental results obtained from \cite{corral2016rosmarinic} as shown in Figs. \ref{Plot_val_B} and \ref{Plot_val_dis}. Fig. \ref{Plot_val_B} shows the normalized mean concentration values of bacterial growth for two different parameter sets. The red dashed line in this figure is obtained via the experimental parameters in \cite{beyenal2003double}. Although this roughly agrees with the growth pattern, the experimental setup in \cite{beyenal2003double} was set for a chemostat which includes a flow of nutrients and dilutes the nutrient concentration to grow in a lower rate. A better fit is obtained by increasing the growth rate and decreasing the nutrient consumption rate as shown with the solid red line in Fig. \ref{Plot_val_B}. In Fig. \ref{Plot_val_dis}, survival percentage of bacteria is shown at the end of $24$ h, which overlaps with the \textit{in vitro} results for the disruption of bacteria.

\begin{figure}[t] 
	\centering
	\includegraphics[width=\columnwidth]{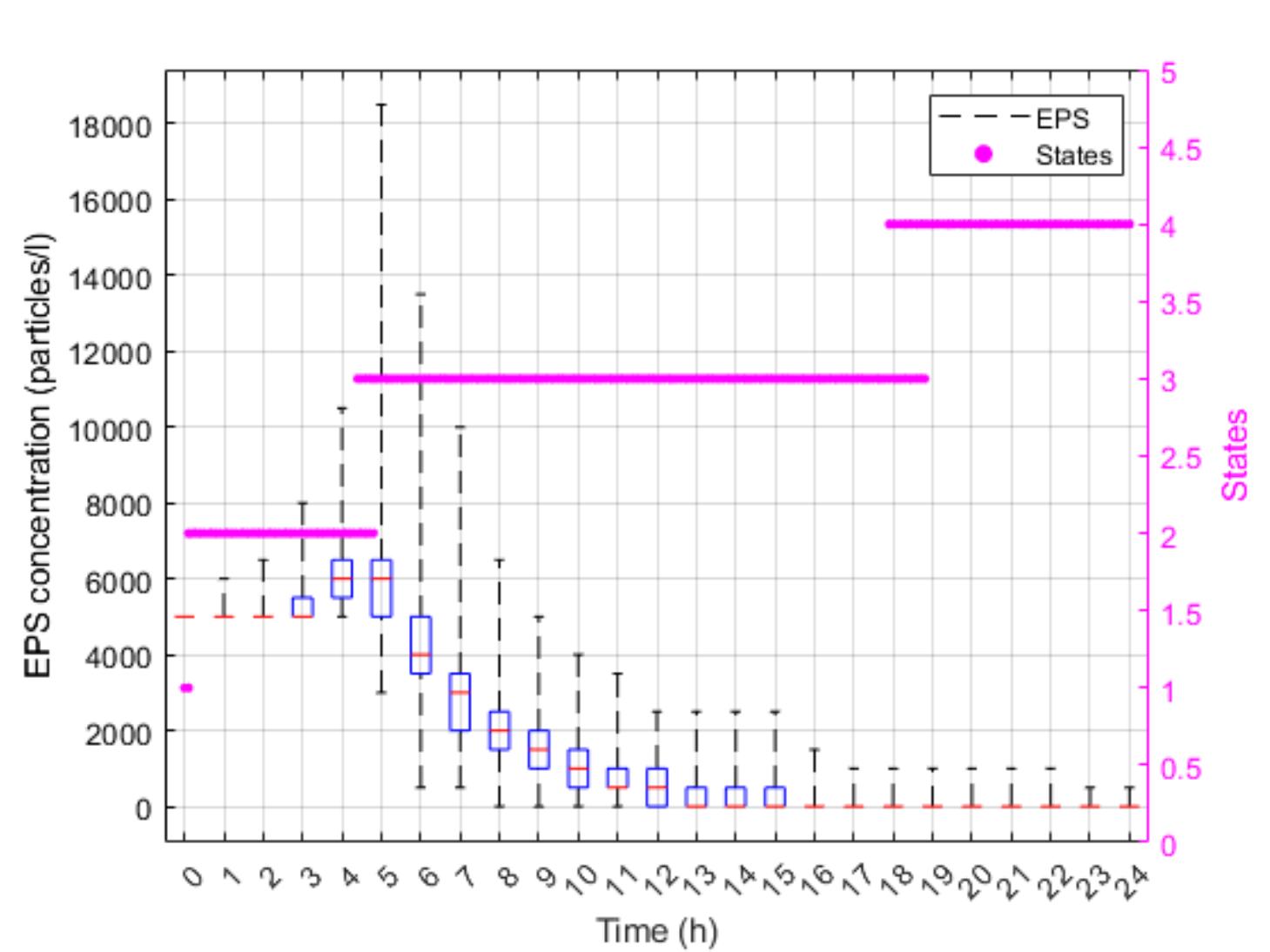}
	\caption{Distribution of states and box plot EPS concentration.}
	\label{Plot_biofilm_E}
\end{figure}

In Fig. \ref{Plot_biofilm_B}, the box plot of the bacterial concentration where the whiskers show the minimum and maximum values, the red line in the box represents the median value and the lower and upper boundaries of the box depict the 25 and 75 percentile, respectively. Furthermore, the distributions for the active times of the biological states stemming from the stochasticity can be observed with respect to the right axis. Although the thresholds are constant for the change of the states, the time for the change of the states can vary as given in the Figs. 4-6 due to the stochasticity of the CRN. Since the growth of bacteria within the biofilm depends only on the nutrient level, bacterial population keeps on growing in states $S_2$ and $S_3$. After nearly $12$ h, the growth stops due to the ending of nutrients. In state $S_4$, the concentration steeply diminishes to zero, since the bacterial disruption only occurs in this state.

\begin{figure}[b] 
	\vspace{-0.5cm}
	\centering
	\includegraphics[width=\columnwidth]{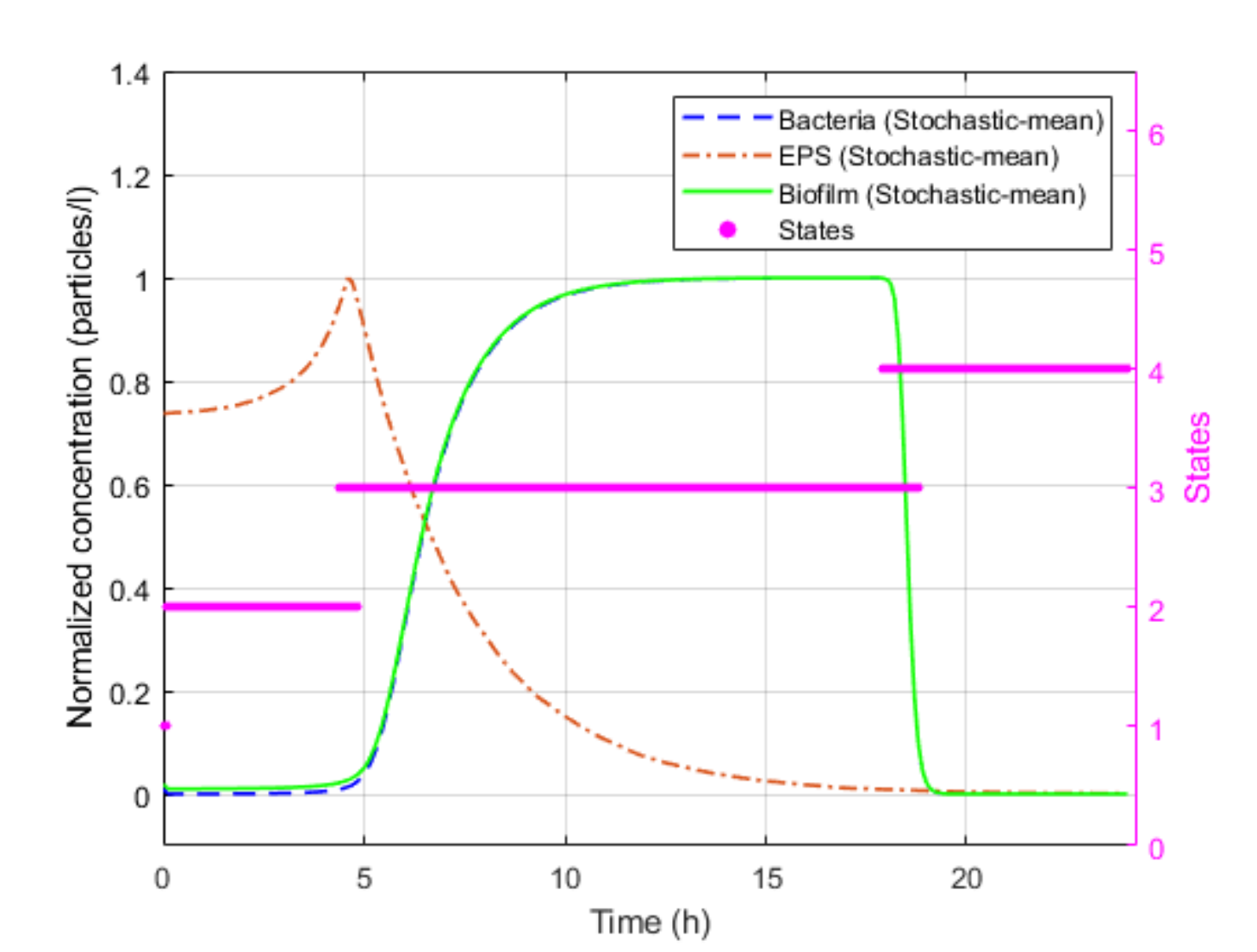}
		\vspace{-0.3cm} 
	\caption{Mean normalized concentration profile of the biofilm.}
	\label{Plot_biofilm_mean}
\vspace{-0.4cm}
\end{figure}

In Fig. \ref{Plot_biofilm_E}, the box plot with the distribution of states is shown for EPS concentration. While bacteria keep on producing EPS, they start to be disrupted by QS mimickers after passing to the third state. Since the EPS disruption rate is much lower than the bacterial disruption rate, the EPS concentration shows a more gradual disruption profile with respect to the bacteria. Moreover, a clearer picture with the mean normalized concentrations of the biofilm and its components (each of the components is normalized according to its local maximum) is depicted in Fig. \ref{Plot_biofilm_mean}. Since the concentration of the biofilm mostly consists of bacteria, the effect of the EPS in the biofilm may seem negligible. This can be observed by the mostly overlapping concentration of biofilm and bacteria in Fig. \ref{Plot_biofilm_mean}. However, the volume covered by the EPS can be larger than bacteria. Moreover, it is observed that there is uncertainty in the transitions between states in Fig. \ref{Plot_biofilm_mean}. For example, it is shown that the system can be in state $S_1$ or $S_2$ at $t=4.5$ h due to the stochasticity of the CRN, although the thresholds for the state transitions are constant. Understanding this uncertainty from a bacterial behavior viewpoint is planned to be researched as the future work.

\section{Conclusion}
In this paper, a stochastic biofilm disruption model based on QS mimickers is proposed. In this model, a CRN is used for the biological processes including QS, production of the biofilm and its disruption. A stochastic state-based simulation algorithm is proposed and results are validated by experimental data. As the future work, the proposed method is planned to be employed to investigate the effect of communication in the bacterial behavior during the biofilm formation and disruption.  

\bibliographystyle{ieeetran}
\bibliography{ref_fg_biofilm_disruption}

\begin{thebibliography}{10}
\providecommand{\url}[1]{#1}
\csname url@samestyle\endcsname
\providecommand{\newblock}{\relax}
\providecommand{\bibinfo}[2]{#2}
\providecommand{\BIBentrySTDinterwordspacing}{\spaceskip=0pt\relax}
\providecommand{\BIBentryALTinterwordstretchfactor}{4}
\providecommand{\BIBentryALTinterwordspacing}{\spaceskip=\fontdimen2\font plus
\BIBentryALTinterwordstretchfactor\fontdimen3\font minus
  \fontdimen4\font\relax}
\providecommand{\BIBforeignlanguage}[2]{{%
\expandafter\ifx\csname l@#1\endcsname\relax
\typeout{** WARNING: IEEEtran.bst: No hyphenation pattern has been}%
\typeout{** loaded for the language `#1'. Using the pattern for}%
\typeout{** the default language instead.}%
\else
\language=\csname l@#1\endcsname
\fi
#2}}
\providecommand{\BIBdecl}{\relax}
\BIBdecl

\bibitem{perez2016mathematical}
J.~P{\'e}rez-Vel{\'a}zquez, M.~G{\"o}lgeli, and R.~Garc{\'\i}a-Contreras,
  ``Mathematical modelling of bacterial quorum sensing: a review,''
  \emph{Bulletin of mathematical biology}, vol.~78, no.~8, pp. 1585--1639,
  2016.

\bibitem{paluch2020prevention}
E.~Paluch, J.~Rewak-Soroczy{\'n}ska, I.~Jedrusik, E.~Mazurkiewicz, and
  K.~Jermakow, ``Prevention of biofilm formation by quorum quenching,''
  \emph{Appl. microbiol. and biotechnol.}, vol. 104, no.~5, pp. 1871--1881,
  2020.

\bibitem{papenfort2016quorum}
K.~Papenfort and B.~L. Bassler, ``Quorum sensing signal--response systems in
  gram-negative bacteria,'' \emph{Nature Reviews Microbiology}, vol.~14, no.~9,
  pp. 576--588, 2016.

\bibitem{corral2016rosmarinic}
A.~Corral-Lugo, A.~Daddaoua, A.~Ortega, M.~Espinosa-Urgel, and T.~Krell,
  ``Rosmarinic acid is a homoserine lactone mimic produced by plants that
  activates a bacterial quorum-sensing regulator,'' \emph{Science Signaling},
  vol.~9, no. 409, pp. ra1--ra1, 2016.

\bibitem{fozard2012inhibition}
J.~A. Fozard, M.~Lees, J.~R. King, and B.~S. Logan, ``Inhibition of quorum
  sensing in a computational biofilm simulation,'' \emph{Biosystems}, vol. 109,
  no.~2, pp. 105--114, 2012.

\bibitem{martins2016using}
D.~P. Martins, M.~T. Barros, and S.~Balasubramaniam, ``Using competing
  bacterial communication to disassemble biofilms,'' in \emph{Proc. of the 3rd
  ACM Int. Conf. on Nanoscale Comput. and Commun.}, 2016, pp. 1--6.

\bibitem{martins2018molecular}
D.~P. Martins, K.~Leetanasaksakul, M.~T. Barros, A.~Thamchaipenet, W.~Donnelly,
  and S.~Balasubramaniam, ``Molecular communications pulse-based jamming model
  for bacterial biofilm suppression,'' \emph{IEEE transactions on
  nanobioscience}, vol.~17, no.~4, pp. 533--542, 2018.

\bibitem{michelusi2016queuing}
N.~Michelusi, J.~Boedicker, M.~Y. El-Naggar, and U.~Mitra, ``Queuing models for
  abstracting interactions in bacterial communities,'' \emph{IEEE J. on Sel.
  Areas in Commun.}, vol.~34, no.~3, pp. 584--599, 2016.

\bibitem{klapper2010mathematical}
I.~Klapper and J.~Dockery, ``Mathematical description of microbial biofilms,''
  \emph{SIAM review}, vol.~52, no.~2, pp. 221--265, 2010.

\bibitem{frederick2011mathematical}
M.~Frederick, C.~Kuttler, B.~Hense, and H.~Eberl, ``A mathematical model of
  quorum sensing regulated eps production in biofilm communities,''
  \emph{Theor. Biol. and Med. Model.}, vol.~8, no.~1, pp. 1--29, 2011.

\bibitem{gillespie1976general}
D.~T. Gillespie, ``A general method for numerically simulating the stochastic
  time evolution of coupled chemical reactions,'' \emph{Journal of
  computational physics}, vol.~22, no.~4, pp. 403--434, 1976.

\bibitem{alvarez2019theoretical}
J.~Alvarez-Ramirez, M.~Meraz, and E.~J. Vernon-Carter, ``A theoretical
  derivation of the monod equation with a kinetics sense,'' \emph{Biochemical
  Engineering Journal}, vol. 150, p. 107305, 2019.

\bibitem{gulec2022stochastic}
F.~Gulec and A.~W. Eckford, ``Stochastic modeling of biofilm formation with
  bacterial quorum sensing,'' \emph{arXiv preprint arXiv:2212.06269}, 2022.

\bibitem{gillespie1977exact}
D.~T. Gillespie, ``Exact stochastic simulation of coupled chemical reactions,''
  \emph{The J. of Phys. Chem.}, vol.~81, no.~25, pp. 2340--2361, 1977.

\bibitem{henkel2013kinetic}
M.~Henkel~et al., ``Kinetic modeling of the time course of n-butyryl-homoserine
  lactone concentration during batch cultivations of pseudomonas aeruginosa
  pao1,'' \emph{Applied microbiology and biotechnology}, vol.~97, no.~17, pp.
  7607--7616, 2013.

\bibitem{beyenal2003double}
H.~Beyenal, S.~N. Chen, and Z.~Lewandowski, ``The double substrate growth
  kinetics of pseudomonas aeruginosa,'' \emph{Enzyme and Microbial Technology},
  vol.~32, no.~1, pp. 92--98, 2003.

\bibitem{gillespie2007stochastic}
D.~T. Gillespie, ``Stochastic simulation of chemical kinetics,'' \emph{Annual
  review of physical chemistry}, vol.~58, no.~1, pp. 35--55, 2007.

\end{thebibliography}

\vfill

\end{document}